\documentclass[aps,pre,showpacs,showkeys,twocolumn,superscriptaddress,
floats,floatfix,preprintnumbers,longbibliography]{revtex4-2}
\usepackage{amssymb,amsmath}
\usepackage[dvipsnames]{xcolor}
\usepackage{bm,url,graphics,subfigure,epsfig,rotating,float}
\usepackage{mathrsfs} 
\usepackage{color,soul,ulem}
\usepackage{comment}
\usepackage{algorithm}
\usepackage{algorithmic}

\graphicspath{{./}{../}{fig/}{../fig/}}
\def \ms {\rm{ms}}

\def \ab    {\alpha\beta}
\def \vd    {\bm{d}}
\def \mI    {\mathcal{I}}
\def \vI    {\bm{\mathcal{I}}}
\def \mIo   {\mathcal{I}_{1}}
\def \mIt   {\mathcal{I}_{2}}
\def \tauA   {\tau_{\rm A}}
\def \DA   {D_{\rm A}}
\def \dx     {{\rm d}x}
\def \dy     {{\rm d}y}
\def \zetaA  {\zeta_{\rm A}}
\def \hhat   {\hat{h}}
\def \Nf     {N_{\rm f}}

\def \qq     {\bm{q}}

\def \grad {{\bm \nabla}}

\def \lap {\nabla^2}

\newcommand{\bra}[1]{\left\langle #1\right\rangle}
\newcommand{\avg}[1]{\left\langle #1\right\rangle}

\def\i{{\rm i}}

\def \mH {\mathcal{H}}

\def \kB {k_{\rm B}}

\newcommand{\delt}[1]{\frac{\partial #1}{\partial t}}
\newcommand{\var}[1]{\text{Var}\left(#1\right)}

\newcommand{\eq}[1]{~(\ref{#1})}
\newcommand{\Eq}[1]{Eq.~(\ref{#1})}

\newcommand{\Fig}[1]{Fig.~(\ref{#1})}
\newcommand{\subfig}[2]{Fig.~(\ref{#1}#2)}

\newcommand{\bfig}{\begin{figure}}
\newcommand{\efig}{\end{figure}}
\newcommand{\bc}{\begin{center}}
\newcommand{\ec}{\end{center}}
\newcommand{\bea}{\begin{eqnarray}}
\newcommand{\eea}{\end{eqnarray}}

\newcommand{\Smat}[1]{(See Supplementary Material)}

\begin{document} 
\title{Estimate of entropy production rate can
   spatiotemporally resolve the active nature of cell flickering}
\author{Sreekanth K Manikandan}
\email{sreekm@stanford.edu}
\affiliation{Department of Chemistry, Stanford University, Stanford, CA, USA 94305
}
\author{Tanmoy Ghosh}
\affiliation{Department of Biological Sciences,
  Indian Institute of Science Education and Research Kolkata,
  Mohanpur, Nadia – 741246, India}
\author{Tithi Mandal}
\affiliation{Department of Biological Sciences,
  Indian Institute of Science Education and Research Kolkata,
  Mohanpur, Nadia – 741246, India}
\author{Arikta Biswas}
\affiliation{Department of Biological Sciences,
  Indian Institute of Science Education and Research Kolkata,
  Mohanpur, Nadia – 741246, India}
\affiliation {Present address: Mechanobiology Institute,
  National University of Singapore, 5A Engineering Drive, Singapore 117411}
\author{Bidisha Sinha}
\email{bidisha.sinha@iiserkol.ac.in}
\affiliation{Department of Biological Sciences,
  Indian Institute of Science Education and Research Kolkata,
  Mohanpur, Nadia – 741246, India}
\author{Dhrubaditya Mitra}
\email{dhrubaditya.mitra@su.se}
\affiliation{ NORDITA, KTH Royal Institute of Technology and
Stockholm University, Roslagstullsbacken 23, 10691 Stockholm, Sweden}

\begin{abstract}
We use the short-time inference scheme (Manikandan, Gupta, and
Krishnamurthy, \textit{Phys. Rev. Lett.} {\bf 124}, 120603, 2020.),
obtained within the framework of stochastic thermodynamics,
to infer a lower--bound to entropy production rate from flickering
data generated by
Interference Reflection Microscopy of HeLa cells.
We can clearly distinguish active cell
membranes from their ATP depleted selves and even 
spatio-temporally resolve 
activity down to the scale of about one $\mu$m.
Our estimate of activity is \textit{model--independent}. 
\end{abstract}
\preprint{NORDITA 2022-034}
\maketitle
\noindent

\section{Introduction}
At the dawn of biophysical research~\cite{schrodinger2012life},
it was already realized that the  fundamental property of living
cells is that they are not in thermal equilibrium
even when they are statistically stationary --
they consume energy and generate entropy, see e.g.,
Ref.~\cite{gnesotto2018broken} for a recent review.
Nevertheless, tools of equilibrium physics are frequently used
to interpret results of experiments on living cells, for example,
equilibrium models are used to infer bending rigidity from
flickering data -- vibrating fluctuations of a cell membrane, see e.g.,
Ref.~\cite{phillips2012physical} for this and several other examples. 
Although the possible influence of active transport on flickering was
first pointed at least seventy years ago~\cite{blowers1951flicker}
investigation into the  essential non-equilibrium feature of
flickering, by comparing flickering data from a healthy cell with its
ATP (adenosine triphosphate)-depleted self, has peaked
in the last two decades~\cite{tuvia1997cell, betz2009atp, rodriguez2015direct},
see also Ref~\cite{turlier2019unveiling} and references therein. 
For RBCs (Red Blood Cells), the first unequivocal demonstration
of their non-equilibrium nature was shown via 
the violation of the fluctuation-dissipation
theorem~\cite{turlier2016equilibrium}.
However, no attempt has been made so far to spatiotemporally resolve,
from flickering data, the
fundamental physical quantity that characterizes non-equilibrium,
v.i.z., the rate of entropy production, $\sigma$.

On the one hand, recent spectacular progress in visualizing and tracking
biological processes~\cite{ntziachristos2010going, brangwynne2008cytoplasmic,
yasuda1998f1, pepperkok2006high, wang2011label}, with unprecedented
accuracy and control raises hopes that measuring entropy production rate for
cellular processes is indeed possible.
On the other hand, there are several difficulties:
(a) They are small in magnitude,  of the order of few $k_B$s per second, and lie below the threshold of detection for
the existing calorimetric techniques~ \cite{basta2018sensitive}.
(b) Thermal fluctuations cannot be ignored hence any measurement results in
noisy readings~\cite{bustamante2005nonequilibrium}.
(c) Typically, we have access to only a few degrees of freedom,
e.g., flickering gives us access to the fluctuations of the cell membrane
down to a certain length and time scale while fluctuations at smaller
length and time scales and the motion of the cytoskeleton that drives
flickering, are not accessible. 
(d) In far-from-equilibrium regimes, there are very few general
principles.

The theoretical understanding of the far-from-equilibrium behavior of
microscopic systems have undergone a revolution over the
last two decades~\cite{jarzynski2011equalities, seifert2012stochastic}
giving rise to the subfield of statistical physics called
\textit{Stochastic thermodynamics}, which can in-principle
extract entropy production rate from long-enough stochastic trajectories
of \textit{all the degrees of freedoms} of the
system~\cite{parrondo2009entropy,seifert2012stochastic}. 
With limited data -- limited in both time and number of accessible
degrees of freedom -- the best we can do is to set
bounds~\cite{esposito2012stochastic,esposito2012erratum,
  barato2015thermodynamic,
  bisker2017hierarchical, lestas2010fundamental, gingrich2017inferring,
  horowitz2020thermodynamic, li2019quantifying, gnesotto2020learning,
  manikandan2020inferring, frishman2020learning,
  otsubo2022estimating, skinner2021improved,roldan2021quantifying,
  martinez2019inferring}. 

In this paper, we use a recent addition to this list of techniques,
\textit{the short-time inference scheme}~\cite{manikandan2020inferring,
  otsubo2020estimating, van2020entropy},
together with flickering data of HeLa cells 
to estimate spatio-temporally resolved entropy production
rate.
As the entropy production rate is the fundamental characteristic of
the active nature of the fluctuations, in the rest of this paper
we shall use the word activity and entropy production rate interchangeably. 

\begin{figure*}
  \includegraphics[width=0.95\textwidth]{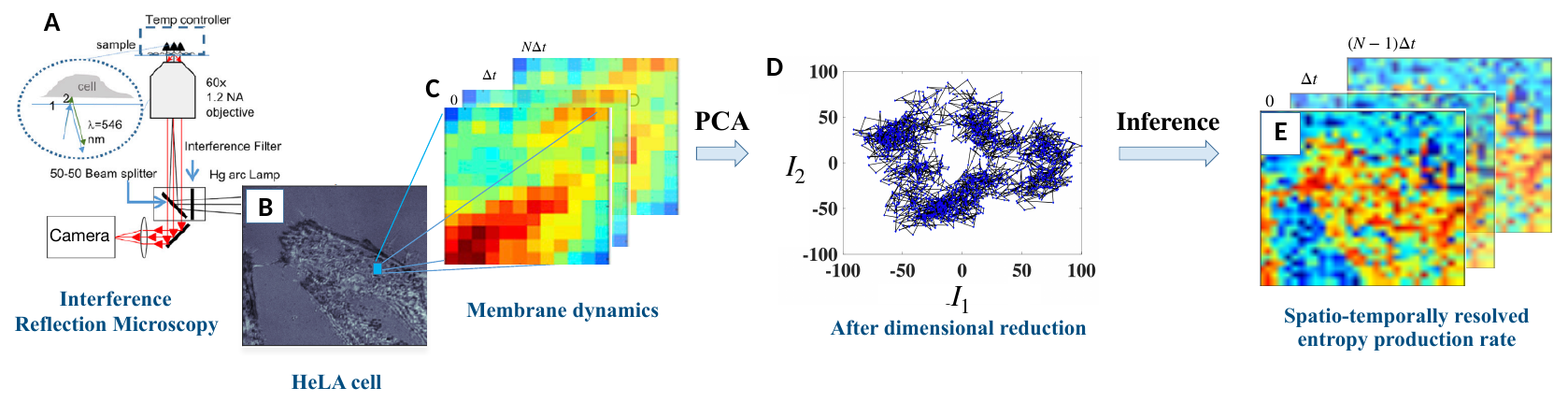}
  \caption{\label{fig:method} We use Interference reflection
    microscopy (A) to obtain flickering data from the basal membrane of HeLa cells (B) with a patch marked by a sky blue square.
    Zoomed in picture of the patch  in (C) typically contains
    $M^2 = 400$ pixels. The $400$ time series are dimensionally reduced to
    a few by Principal Component Analysis (PCA)
    (D) The stochastic trajectory of the first two principal components
    for a typical case.  We obtain the entropy production rate, $\sigma$
    by numerical optimization of this reduced problem. Using the
    first six principal components instead of the first two
    has a negligible effect on our results.
    Color map of $\sigma$ in (E) clearly shows the active cells
    compared to its background. The sketch in (A) is adapted
  from Ref.~\cite{biswas2017mapping}.} 
\end{figure*}
The short-time inference scheme is built on the thermodynamic uncertainty
relation~\cite{barato2015thermodynamic}
which states that a lower bound to the entropy production rate
can be obtained via fluctuations of any arbitrary current $J$ in phase space,
as,
\begin{equation}
\label{eq:seifert}
    \sigma \geq 2\kB \frac{\bra{J}^2}{t\;\var{J}}.
\end{equation}
where $\bra{\cdot}$, and $\var{\cdot}$ denotes mean and variance of a random
variable calculated from a statistically stationary time-series of 
phase space trajectories over a time-interval  $t$ and $\kB$ is the
Boltzmann constant. 
Refs.~\cite{manikandan2020inferring, otsubo2020estimating, van2020entropy}
extended this result to show that 
for a large class of non-equilibrium systems with an overdamped Langevin
dynamics, the inequality, \eq{eq:seifert}, saturates in
the short time limit, i.e., 
\begin{align}
\label{eq:stis}
\sigma = 2\kB\lim_{\Delta t \to 0}\max_J 
\left[\frac{\bra{J_{\Delta t}}^2}{\Delta t\;\var{J_{\Delta t}}}\right]\/.
\end{align}
Here $J_{\Delta t}$  is calculated over time $\Delta t$ and the mean
and the variance is calculated over an ensemble of such time intervals.

This scheme has certain key advantages over its competitors:
one, it is model independent -- the precise knowledge of the underlying
overdamped Langevin equation is not required;
two, it is not limited to stationary time series.
This scheme has been successfully used to estimate the entropy production rate
from a moderate number of realizations of stochastic trajectories of
time-dependent systems~\cite{otsubo2022estimating}. 
The scheme was also tested experimentally  using the stationary
trajectories of a colloidal particle in an
stochastically--shaken optical trap~\cite{manikandan2021quantitative}. Here we apply \Eq{eq:stis} to analyze flickering data from
living cell membranes.

\section{Methods}

In \Fig{fig:method} we pictorially summarize our method. 
We first obtain a movie of flickering data for the basal membrane, a part of the membrane attached to the substrate, of adherent HeLa cells using Interference Reflection Microscopy (IRM). HeLa cells (CCL-2, ATCC) were grown in Dulbecco’s Modified Essential Medium (DMEM, Gibco, Life Technologies, USA) supplemented with $10\%$ fetal bovine serum (FBS, Gibco) and $1\%$ Anti-Anti (Gibco) at $37^\circ \mathrm{C}$ under $5\%$ $\mathrm{CO}_2$. Experiments were always performed after 16-18 hours of cell seeding. IRM imaging \cite{biswas2017mapping,limozin2009quantitative} was conducted using a motorized inverted microscope (Nikon, Japan) equipped with adjustable field and aperture diaphragms, a $60\times$ water immersion objective (NA 1.22), $1.5\times$ external magnification, an onstage $37^\circ \mathrm{C}$ incubator (Tokai Hit, Japan), an s-CMOS camera (ORCA Flash 4.0, Hamamatsu, Japan), a $100$ W mercury arc lamp, an interference filter ($546 \pm 12$ nm), and a 50-50 beam splitter \cite{biswas2017mapping}. Pixel intensities were converted to relative heights (distance of the membrane from the coverslip) using a MATLAB (MathWorks, USA) code after obtaining the intensity-to-height conversion factor determined by a beads-based calibration method as reported previously \cite{biswas2017mapping}. Every experiment was preceded by an independent calibration.

A typical image  of a cell is shown in \subfig{fig:method}{B}.
In addition to the cell the background is also visible.
For the analysis, we divide this cell into several square patches of 
equal size ($\sim 2 \mu m^2$).
One such patch is shown as a blue square in \subfig{fig:method}{B}.
Each patch is made of $M\times M$ pixels.
Thus the time evolution of each patch is completely described
by $M^2$ stochastic variables $I = (I_1, I_2, \dots, I_{M^2})$ -- 
the light intensity at each of the pixels.
We record movies with a time-step $\Delta t = 50\/\ms$ between
two consecutive snapshots. A typical movie consists of $\Nf = 2048$
snapshots - see \subfig{fig:method}{C}. 
In order to apply the inference scheme in \eq{eq:stis}, in principle, we can
calculate currents from two consecutive
snapshots and average over the number of snapshots to calculate
the quantity inside the square brackets in \eq{eq:stis}.
In that case, we have to perform numerical optimization over
$M^2$ space which is a formidable problem.
Following Ref.~\cite{gnesotto2020learning} we use
Principal Component Analysis (PCA) to reduce the dimension of the
problem in the following manner.
Construct the covariance matrix ${\bm C}$, with elements 
$C_{ij} \equiv \bra{I_i I_j} - \bra{I_i}\bra{I_j}$. 
The average is taken over the number of frames, $\Nf$. 
This is averaging over a time interval, in which the time series 
is assumed to be stationary.
The method applies equally well for ensemble averaging~\cite{otsubo2022estimating}. 
The projections of the data along the eigenvectors of this covariance 
matrix, in decreasing magnitude of eigenvalue, are the principal components.
\subfig{fig:method}{D} shows a typical example of the dynamics in the first two components, 
$\mIo(t)$ and $\mIt(t)$.
In general, the state of the system can be projected to any $N \leq M^2$ dimensional vector 
$\vI = (\mI_1,\ldots \mI_{\alpha},\ldots \mI_N)$. 
Next, we define a scalar current:
\begin{equation}
  J_{\Delta t}(t_i) \equiv \vd\left(\frac{\vI(t_i+\Delta t)+\vI(t_i)}{2} \right)\cdot
  \left(\vI(t_i+\Delta t)-\vI(t_i)\right),
  \label{eq:J}
\end{equation}
where $\vd(\mI)$ is \textit{any} arbitrary $N$ dimensional function. 
Then we need to find the optimal function $\vd^*(\mI)$ which maximize the bound in \eq{eq:stis}. 
We use two different algorithms:  
(A) We use a linear function $d_{\alpha}(\vI) = c_{\ab}\mI_{\beta}$,
where the matrix $c_{\ab}$ is a matrix of constant coefficients and
perform the optimisation by the particle--swarm algorithm.  
(B) We use a non-linear function for $\vd(\mI)$ represented by a feed--forward 
neural network 
$\vd(\mI\vert {\bm \theta})$, 
where $\theta$ are the parameters of the network. 
For a fixed choice of the parameters, the mean and the variance of the current are
computed by averaging over the index $i$.
The analysis can then be extended to cells under different physiological conditions and 
at different times, to obtain a spatio-temporally resolved entropy production map as in 
\subfig{fig:method}{E}.
Both the algorithms give similar results; with (A) yielding a slighly lower
estimate than (B). See Appendix \ref{sec:InferenceAlgorithms} for details.

\section{Results}

\begin{figure*}
  \includegraphics[width=0.95\textwidth]{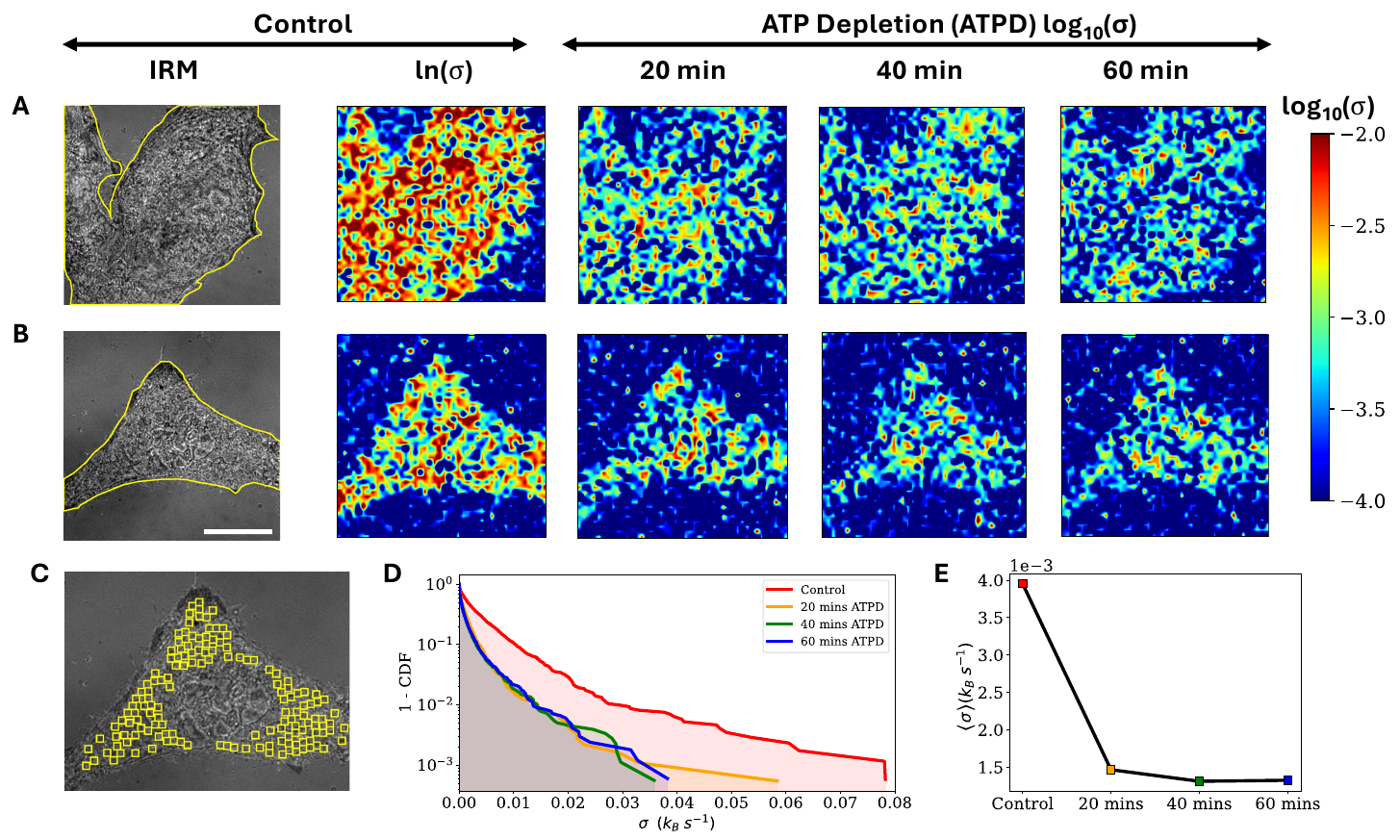}
	\caption{\label{fig:atpd} \textbf{Activity of cell membrane}.
    (A) and (B) From left to right: Typical IRM image of two live HeLa cells
    -- the boundary of the cells is marked in yellow.
    Color map of  $\ln(\sigma)$ before the cells are treated with ATP-depleting
    agents. Color map of $\log_{10}(\sigma)$ after incubating the cells in
    ATP-depleting agents for $20$min., $40$min. and $60$min.
    For clarity, the color map is limited to a fixed range. 
    (C) Representative IRM image of a cell with FBRs
    (size: $20\times 20$ pixels or $2 \mu {\rm m}^2$) marked
    in yellow
(D) Plot of (1 - CDF) of $\sigma$, for the FBR--patches, for control and three different 
durations of ATP-depletion. 
 (E) Average of $ \sigma$ over FBR patches  under ATP-depletion. }
\end{figure*}

In \subfig{fig:atpd}{A}, we show how our method performs on the experimental
data as the cells are ATP-depleted by incubating them in glucose-free ATP-depleting medium. 
We do this by adding 10 mM sodium azide and 10 mM 2-deoxy D-glucose to the cells \cite{renard2015endophilin} in M1 Imaging medium (150 mM NaCl; Sigma-Aldrich), 1 mM MgCl2 (Merck, Kenilworth, NJ), and 20 mM HEPES (Sigma-Aldrich), and incubating them for 60 minutes for ATP depletion.
In recent work by some of the authors, it was found that ATP-driven activities increase temporal fluctuations and 
flatten out spatial undulations --- see Figure 3 in \cite{biswas2017mapping} and discussions therein for the detailed analysis and discussion on the effect of ATP depletion on membrane fluctuations.

In the leftmost row of \Fig{fig:atpd} we show the image of the cell -- the cell boundaries
are marked in yellow. 
The next column shows a pseudocolor plot of $\log_{10}(\sigma)$
of the live cell --  marked by \textit{control}. 
The next three columns show how the entropy production rate changes
after $20$ minutes, $40$ minutes and $1$ hr of incubation.
In each of these cases, we use a movie of $\Nf = 2048$ snapshots where
consecutive snapshots are separated by $\Delta t = 50\rm{ms}$.
Since the total duration of the movie is quite short compared to the
time-scales over which ATP-depletion operates, each of these
movies are considered statistically stationary.
We have performed these experiments for a total $31$ HeLa cells,
with two independent repeats. 
Images of two typical cells are shown in \subfig{fig:atpd}{A} and \subfig{fig:atpd}{B}.
The images of several other representative cases are in the supplemental material.
Our results clearly demonstrate that the entropy production rate computed
using \eq{eq:stis} spatiotemporally resolves the active nature of cell membrane
fluctuations.
Notably, we see many patches of high entropy production in the
active cell membrane which is well contrasted with the background,
and patches of low entropy production in the
ATP--depleted membrane, less contrasted with the background.
We also see an
overall decay of $\sigma$ in time.

To calculate reliable statistical properties of the entropy production rate
-- for the control and the ATP-depleted ones -- we limit ourselves to patches
that are inside the cell and away from the nucleus -- an example is
shown in \subfig{fig:atpd}{C}.
The relative heights of these patches lie within the range 
$\sim 0-100$ nm -- termed as first-branch regions (FBRs).
Henceforth we call these patches FBR--patches. 
We use rank--order methods to calculate the cumulative distribution function (CDF) of
$\sigma$ from all the FBR--patches of all the cells, see \subfig{fig:atpd}{D}.
The rank order method does not suffer from the binning errors that plague
histogram based methods to calculate the probability distribution function. 
We calculate the PDF of $\sigma$ by calculating the derivatives of the CDF.  
The complement of the CDF, defined as 1 - CDF has an exponential tail to the right. 
This implies that the PDF of $\sigma$, also has an exponential tail. 
In \subfig{fig:atpd}{E} we show how the mean value of the $\sigma$
decreases under ATP depletion.   
A straightforward computation of the average $\sigma$ over all FBR patches for the control 
cell yields a value of approximately $4 \times 10^{-3} \, \kB \, s^{-1}$ across a patch 
area of $2 \mu m^2$.
The exponential tail of the PDF implies that $\sigma$ shows large fluctuations.  
In particular, the average of the top 10\% of patches is nearly five times higher 
than the mean, and the average of the top 50\% of patches is nearly 
double --- See Appendix \ref{sec:percentile}.

\begin{figure*}
\includegraphics[width=0.95\textwidth]{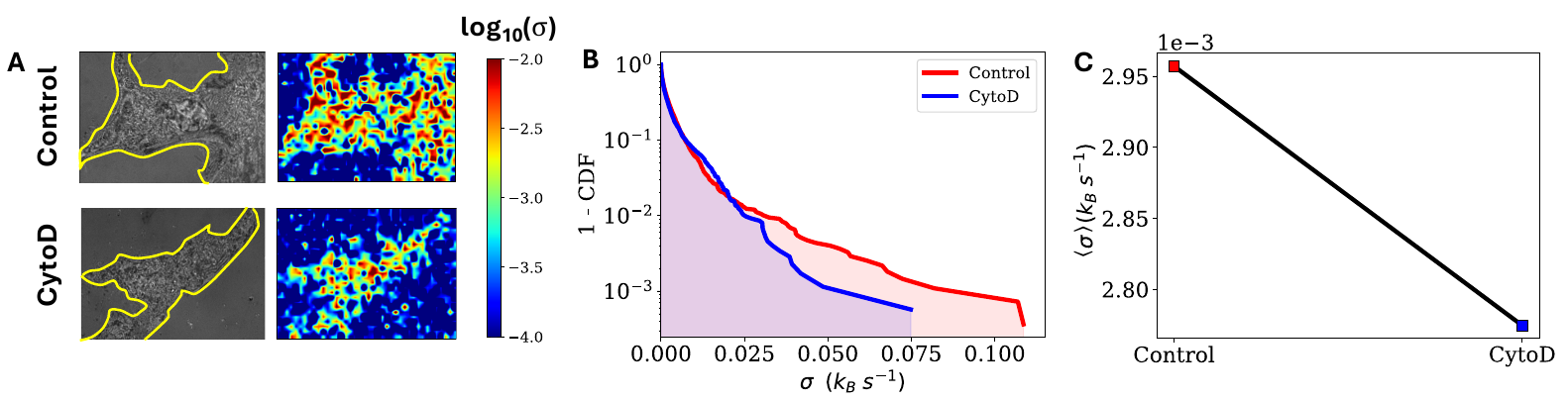}
\caption{\label{fig:cytoD} \textbf{How Cyto-D affects activity?}
(A) IRM image and colormap of $\log_{10}(\sigma)$ of a HeLa cell from the control set and Cyto D
  treated cells.
Note that unlike \Fig{fig:atpd} these two are not the same cell.
 The boundary of the cells are marked in yellow.
(B) The average as well as the distribution (inset) of $\sigma$. 
Data is representative of three independent experiments
on $17$ control cells and Cyto-D treated ones. 
The images of several other representative cases are in the supplemental material.
}
\end{figure*}

Which processes on the membrane are the major contributors to the measured
activity?
Of the many ATP-dependent processes that may actively impact membrane
fluctuations, forces originating at the underlying actomyosin cortex are
important~\cite{koster2016cortical, biswas2017mapping}.
The effect of actin polymerizing forces~\cite{dmitrieff2016amplification}
as well as forces by myosin motor protein contracting the actin network
are expected to be diminished if the cortex is weakened or removed.
It has been shown~\cite{schliwa1982action} that the drug
Cytochalasin D (Cyto D) suppresses actin polymerization
and weakens the cortical network. This is achieved by treating the cells with $5$ $\mu M$ cytochalasin D (Cyto D, Sigma Aldrich) for 1 hour in serum-free media \cite{biswas2017mapping}.
It's effect on membrane fluctuations was further found to be highly spatially inhomogeneous --- see Figure 4 in Ref.\ \cite{biswas2017mapping} and discussions therein for the detailed analysis and discussion on the effect of Cyto-D treatment on membrane fluctuations.
The network is  subsequently contracted and fragmented by myosin,
the cortex ruptures, clears from most of the membrane area and accumulates 
at multiple foci.
Thus we expect that a possible effect of cyto-D is to decrease
the entropy production rate.
In \Fig{fig:cytoD} we show the effect of cyto-D on the entropy production rate. 
We indeed find a decrease in activity.
The decrease of the mean value of the measured activity is smaller than the decrease measured for ATP depletion.

Note that the probability distribition function of entropy generation 
rate has an exponential tail. 
We have found a significant number of patches with estimates 
of $\sigma$ at least $5$ times higher than the mean value 
(see Fig.~5 of supplemental material). 
As our measurements find set a lower bound to the actual 
entropy production rate, it is likely that in reality the
PDF falls off even slower than an exponential.
This underscores the heterogeneity nature of  activity  
of the membrane.  

\section{Discussion}

Our attempt to tease out the essential non-equilibrium feature from 
flickering by using the entropy production rate as a measure of
activity has a clear advantage over earlier methods, such as the ones
dependent on the breakdown of the fluctuation-dissipation relation e.g.
Refs.~\cite{turlier2016equilibrium,ariga2018nonequilibrium,
  ariga2020experimental,ariga2}.
First, the entropy production rate is the crucial measure of non-equilibrium --
if a system A has a higher entropy production rate than a system B
then A is  \textit{further away} from equilibrium than B.
Second, we are able to spatiotemporally resolve the entropy production rate
thereby able to identify regions that have higher entropy production
rate. This has not been achieved before.
Third, our method gives a model-independent estimate of the lower-bound
of entropy production whereas 
Refs.~\cite{turlier2016equilibrium,di2024variance} must use a model for active
fluctuations.
Finally, the measurements in Refs.~\cite{turlier2016equilibrium,
  ariga2018nonequilibrium, ariga2020experimental,ariga2, di2024variance} are 
typically invasive in nature -- they involve tracking the response of 
microscopic beads attached to systems and external 
perturbations~\cite{brangwynne2008cytoplasmic, brangwynne2008cytoplasmic} or 
attaching fluorescent proteins or filaments to
the relevant degrees of freedom~\cite{yasuda1998f1,pepperkok2006high}.
In contrast, our analysis is less invasive and is applicable to movies made by
exploiting the naturally existing intracellular contrast~\cite{wang2011label,
  gnesotto2020learning}.

Several comments are now in order.
First, on the role on noise in our data. 
Clearly, if we apply our algorithm to patches outside the cell 
we would obtain a small but nonzero number. 
We define a signal--to--noise ratio to be the  activity obtained for 
patches inside the cell over patches outside the cell.
We calculate this for patches of different sized and find that the
signal--to--noise ratio is maximum for patches of size $2 \mu m^2$
(Fig 3A in supplementary material).
All the results we show in this paper uses this size. 
Furthermore, for this patch size the activity grows linearly with
area of the membrane. See Appendix \ref{sec:Area_Scaling}.

Second, in addition to the necessary spatio-temporal coarse--graining,
which is part of any experimental measurement,
we  have performed a dimensional reduction by  using 
principal component analysis and retained the first few components.  
How many principal components must  we keep?
We use a method  proposed in Ref.~\cite{gnesotto2020learning}.
We randomly shuffle our data along the time axis and then again
calculatate the eigenvalues of the correlation matrix.
We find that the eigenvalues of this shuffled data
are very close to one another.
These set the noise floor of our measurements of principal components.
Now going back to the eigenvalues calculated from the original data
we find that only the first few, at most $10$, eigenvalues are 
significantly larger than the noise floor (see Appendix \ref{sec:PCA}).
Next, let us consider the algorithm where  we assumed the
optimal function to be a linear function of the currents. 
We limit our analysis to the first two principal components. 
Increasing the number of components to $10$ do not have any significant 
effect on the estimated value of $\sigma$. 
For the second algorithm, where we used a  neural network representation,
the estimate of $\sigma$ changes as the number of components are increased
from $2$ to $10$, \textit{during the training stage}. 
But yielded minimal change in the \textit{testing stage}. 
Nonetheless, we retained the first $10$ components for all the results
shown in this paper.  
The estimates of $\sigma$ obtained by the two algorithms 
are comparable; the neural network-based estimates are slightly higher.
Furthermore, these also have better signal--to--noise ratio between the cell 
and the background.

Third, we compare our findings with the already known estimates of
entropy production rates in microscopic biological systems.
To the best of our knowledge there are no results for HeLA cells. 
Measurements that tracked a single filament of microtubules report an
entropy production rate of $5 \kB\rm{min}^{-1}$~\cite{skinner2021improved}
while a single actin fiber contracted by myosin motors report
an entropy production rate, per unit length, 
to be $\sim$ 1 $\kB/(\mu\rm{m}s)$~\cite{seara2018entropy}.
At the level of cells, e.g., in  Ref.~\cite{roldan2021quantifying},
the entropy production rate was
quantified from experimental recordings of spontaneous hair-bundle
oscillations in mechanosensory hair cells from the ear of the bullfrog.
Interestingly, by applying \eq{eq:seifert}, without any optimization
over the current, the study found entropy productions of the order
of $10^3 \kB s^{-1}$. 
Another recent work~\citep{di2024variance} that analyzes
flickering data from circumference of RBCs in a model dependent fashion 
finds a entropy production rate of $\approx 10^6 \kB s^{-1}$.  
This estimate depends on the details of the model used for cell-flickering. However, it is argued to be compatible with microcalorimetric measurements of heat generation from packed RBCs 
in bulk~\cite{bandmann1975clinical,backman1992microcalorimetric}.
In contrast, we find an average entropy production rate of   
$\sim 4\times 10^{-3}\; \kB {\rm s}^{-1}$ over a $2 \mu m^2$
patch of the membrane. The distribution of the entropy production rate over patches feature a fat tail to the right, and there are a significant number of patches with activities up to five times higher than the mean value {\color{black}(see Appendix \ref{sec:percentile})}.
Multiplying just the mean with the area of the basal membrane ($\sim 10^3 \mu m^2$) 
we find an estimate close to  a few $\kB$s per second.
Although, this is at least an order of magnitude higher than the numbers reported 
in Ref.~\cite{skinner2021improved}, who used measurements of
calcium flicker trajectories from single cells
reported in Ref.~\cite{thurley2014reliable}, 
these values are several orders of magnitude lower than 
Refs.~\citep{roldan2021quantifying, di2024variance}. 
 There may be several reasons for the low estimates we obtain, but the obvious reasons appear to be high noise levels in the experimental data and lower bounds of the estimate in Eq.\ \eqref{eq:stis} resulting from the necessary spatiotemporal coarse-graining and dimensional reduction. 
Furthermore, we reiterate that the goal of our paper is not to calculate the total entropy
production by all non-equilibrium processes in a cell
-- our focus is solely on the cell membrane.
We do expect values significantly smaller than the total
rate of consumption of ATPs in a cell because 
the membrane fluctuations are just one non-equilibrium process 
among many that goes on in a living cell.

Fourth, can we estimate the time-scales of the active processes that play the
most significant role in entropy generation?
Unlike many of the cases listed above,
the flickering data does not contain any clear ``active spikes''.
This may be because multiple active processes with a range
of time scales, (0.1-2s~\cite{turlier2016equilibrium,biswas2017mapping})
act on the membrane.
Furthermore, an estimate of time--scales  is necessarily model-dependent. 
We have analyzed such a model to obtain a closed form estimate for the 
entropy production rate, and argue why it cannot be reliably
applied to experimental data. See Appendix \ref{sec:Amodel} for details.

In summary, there are a couple of areas that could benefit from further studies. Firstly, the entropy generation rate we obtained may be too small to be relevant for the metabolic processes of the whole cell. This could be attributed to the high noise level in our data; at most, only the first ten, sometimes even just the first two, principal components contain useful information. Thus, we anticipate that with improved resolution, our approach may unveil a significantly larger entropy generation rate. Secondly, the effect of Cyto-D on the entropy generation rate is quite small. Despite the significant impact of the acto-myosin cytoskeleton, which is notably affected by Cyto-D, the most active fluctuations of the membrane remain largely unchanged. Consequently, although we expected Cyto-D to substantially alter the entropy production rate, this is not the case. At present, we do not fully comprehend this aspect of our results.

To end on a positive note, 
besides capturing the cell--level activity,
we report for the first time, a spatiotemporally resolved activity. 
As shown in Fig.\ \ref{fig:atpd}, cells have high activity regions which disappear on
ATP depletion. This offers explicit evidence of heterogeneity of activity in the cell
membrane pointing to a lack of rapid establishment of an
statistically homogeneous (although non-equilibrium) state. 
This technique can be combined with fluorescence imaging of a variety of
``active'' proteins or structures to measure their localized action.
Thus, comparing maps of lateral organization of molecules of interest with
activity maps could potentially identify those involved in actively regulating
processes like cell migration, endocytosis, and mechano-sensing.

\acknowledgements
SKM acknowledges the Knut and Alice Wallenberg Foundation for financial support through Grant No. KAW 2021.0328.
NORDITA is partially supported by Nordforsk.
DM acknowledges the support of the Swedish Research Council
Grant No. 638-2013-9243 and 2016-05225.
BS acknowledges support from Wellcome Trust/DBT India Alliance
fellowship (grant number IA/I/13/1/500885) and CSRP from
CEFIPRA (grant number 6303-1). TG thanks CEFIPRA for scholarship. 
TM is supported by a predoctoral fellowship from the Council of
Scientific and Industrial Research (CSIR), India. 

\appendix
\section{Inference algorithms}
\label{sec:InferenceAlgorithms}
We first considered a particle swarm-based algorithm to perform the maximization:
\begin{align}
\label{eq:Astis}
\sigma = 2\kB \max_J
\left[\frac{\bra{J_{\Delta t}}^2}{\Delta t\var{J_{\Delta t}}}\right].
\end{align}

Here, $J_{\Delta t}$ is calculated over time $\Delta t$, and the mean and the variance are calculated over an ensemble of such time intervals. Explicitly, we are searching over the phase space of the first few PCA components (see subsection \ref{sec:PCA}), and currents constructed using a linear basis. We defined an $N$-dimensional vector $\vd(\vI)$ as a linear combination of the components of $\vI$, i.e.,

\begin{equation}
d_{\alpha}(\vI) = c_{\alpha\beta}\mathcal{I}_{\beta},
\label{eq:d}
\end{equation}

where the matrix $c_{\alpha\beta}$ consists of constant coefficients. At time $t=i\Delta t$, we define a scalar current:

\begin{equation}
J_{\Delta t}(t_i) \equiv \vd\left(\frac{\vI(t_i+\Delta t)+\vI(t_i)}{2} \right)\cdot
  \left(\vI(t_i+\Delta t)-\vI(t_i)\right).
\label{eq:J}
\end{equation}
While using the particle swarm algorithm, we fixed $N=2$. Increasing the number of components was did not have a significant effect on the estimates of entropy production rate, primarily due to the presence of experimental noise. The details of the algorithm can be found in Ref. \cite{manikandan2021quantitative}.

Linear representations of the function $\vd(\vI)$ such as the above severely limits the phase space of currents we are working with. Ideally, we should consider models which are more generic in the sense of having a non-linear basis, and straightforwardly scalable to arbitrary number of principle components. The most natural choice is to use a Neural Network. Hence we have performed and independent estimation of the entropy production rate by using a neural network to represent the vector $\vd(\vI)$. The network architecture we used  consists of two fully connected layers with a hyperbolic tangent (Tanh) activation function applied after each layer. The input dimension ($\texttt{x\_dim}$) is set to $N$, and the number of nodes in the hidden layer ($\texttt{nb\_nodes}$) is set to 256. For any input, this network returns an $N$ dimensional, non-linear representation of the vector $\vd := \vd(\mI \vert {\bm \theta})$,
where ${\bm \theta}$ are the parameters of the nerual network.
The presence of the activation function $\tanh$ makes this representation arbitrarily non-linear, and bounded.

\begin{figure*}
    \centering
    \includegraphics[width = \linewidth]{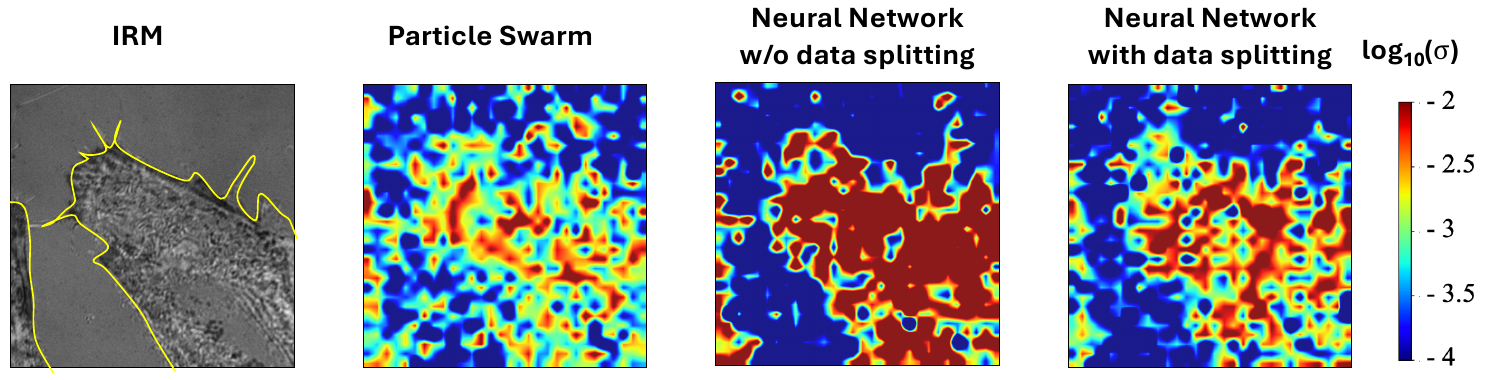}
    \caption{The figure illustrates comparisons of entropy production rate estimates obtained using different algorithms: Particle swarm algorithm which takes in as input the first two principle components of IRM data, and neural network based algorithm which takes as input the first 10 principle components. While the neural network-based algorithm, without a data-splitting scheme, produced the highest estimate, discerning overfitting effects is challenging. Conversely, estimates acquired using neural networks following the data-splitting scheme were lower but comparable in order of magnitude and slightly higher in values to those obtained from the particle swarm algorithm. Thus, we have presented results only from the neural network with data splitting and the particle swarm algorithm.}
    \label{fig:inferencecomparison}
\end{figure*}

In Algorithm 1, we provide a schematic summary of the inference scheme with the neural network model. The hyperparameters of the algorithm, such as the number of nodes in the hidden layer, the number of Principle components used and the number of steps used for training are chosen by trying to maximize the signal-to-noise ratio between the estimate of entropy production rate between the cell and the background. Crucially, we have also used a data-splitting scheme where the first half of the data is used for training the model and the second half of the data is used to estimate entropy production rate using the trained model. See \cite{otsubo2022estimating} for details. This approach helps overcome issues of overfitting to the data, which can lead to very high estimates of dissipation. Overfitting was also found to happen when the training was done for a large number of steps. Hence, we limited the number of steps in training to 100. With the data-splitting scheme, we get estimates of dissipation that is consistent with the estimates obtained using the Particle Swarm algorithm. See Fig.\ \ref{fig:inferencecomparison} for the comparison of results between the two algorithms. Results for more cells using both algorithms are shown in the supplemental material.

\subsection{Dependence of the entropy estimate on the patch size}
\label{sec:Area_Scaling}
\begin{figure*}
    \centering
    \includegraphics[width=0.9\linewidth]{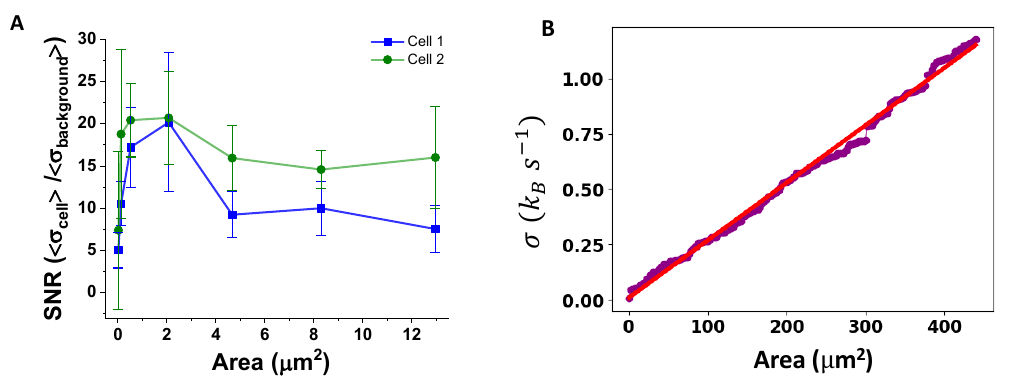}
    \caption{(A) The signal-to-noise-ratio computed by comparing the ratio of
      activity inferred in FBR region with the corresponding value for the
      background.
      We find that the  patch size $20 \times 20$ pixels
      or $\sim 2\; \mu\; m^2$ has the best signal to noise ratio.
      (B) The dependence of the entropy production rate on the area of the
      cell-membrane for the optimal patch size ( $20 \times 20$ pixels ).
      We observe that the estimate increases approximately linearly with the
      area.  }
    \label{fig:Area_Scaling}
\end{figure*}

In Fig.\ \ref{fig:Area_Scaling}, we summarise our analysis of how entropy
production estimates depend on the cell area.
We identified an optimal patch size of $20 \times 20$ pixels
(approximately $2; \mu; m^2$), which yielded the highest signal-to-noise ratio
when comparing activity within the FBR region to background values. For this optimal patch size, we further found that entropy production computed over the entire cell membrane increases linearly with the membrane's area (number of patches). 

\subsection{Principle Component Analysis}
\label{sec:PCA}
\begin{figure}
    \centering
    \includegraphics[width = 0.9\linewidth]{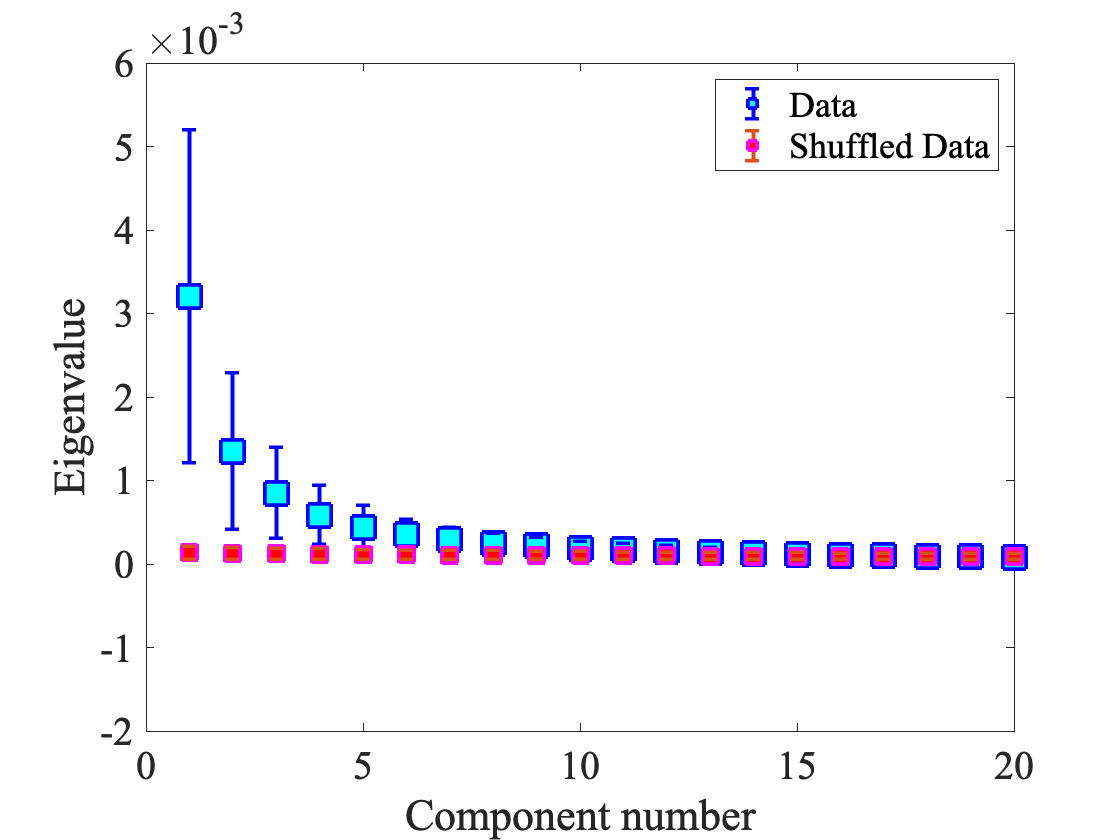}
    \caption{Plots of the first 20 eigenvalues corresponding to PCA components,
      for the normalized experimental data (blue) and the normalized data
      randomly shuffled along the time axis (purple), for a representative cell flickering data.
      The errorbars correspond to one standard deviation over different
      patches.
      The PCA eigenvalues for data shuffled along the time axis shows the
      noise level in the system~\cite{gnesotto2020learning},
      which lie slightly above zero.
      We see that the magnitude of the eigenvalues rapidly drop and reach the
      noise levels within the first few principle components.}
    \label{fig:PCA_Error}
\end{figure}

Here we demonstrate that the principle components other than the first few do not significantly contribute to the dynamics. We consider the data corresponding to the first cell in Figure 2 of the maintext. 
In Fig. \ref{fig:PCA_Error}, we plot the first 20 eigenvalues corresponding to PCA components for the normalized experimental data (Blue) and the normalized data randomly shuffled along the time axis (Purple). The error bars represent the standard deviation over different patches. The PCA eigenvalues for the shuffled data show the noise level in the system \cite{gnesotto2020learning}, which is slightly above zero. We observe that the eigenvalues rapidly decrease in magnitude and reach the noise levels within the first few principal components. This indicates that the principle components other than the first few closely approach the noise floor and do not significantly contribute to the dynamics.

\subsection{Percentile Averages}
Due to significant inhomogeneity across the cell surface, relying solely on the mean value of the entropy production rate over patches may be inadequate. Hence here we look at percentile averages, where we plot the average $\sigma$ over the top $x \%$ of patches of the control cells.  As shown, the average of the top 10\% of patches is nearly five times higher than the mean, and the average of the top 50\% of patches is nearly double. 
\label{sec:percentile}
\begin{figure}
    \centering
    \includegraphics[width=0.9\linewidth]{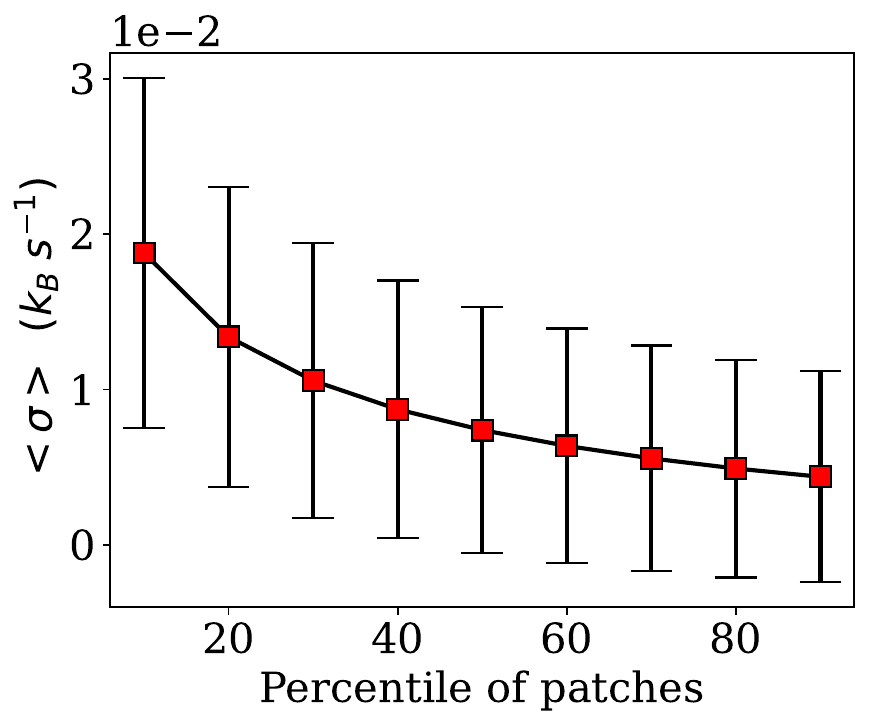}
    \caption{The average $\sigma$ over the top $x \%$ of patches of the control cells. Error bar corresponds to the standard deviation. We observe that the average of the top 10\% of patches is nearly five times higher than the mean, and the average of the top 50\% of patches is nearly double.
    Data is from one set of experiments using $14$ cells.
    Two independent repeats show a similar trend. }
    \label{fig:percentile}
\end{figure}

\section{Entropy production rate in a model of active membrane}
\begin{figure*}
\label{fig:sigmaqplot}
\centering
\includegraphics[width=.9\linewidth]{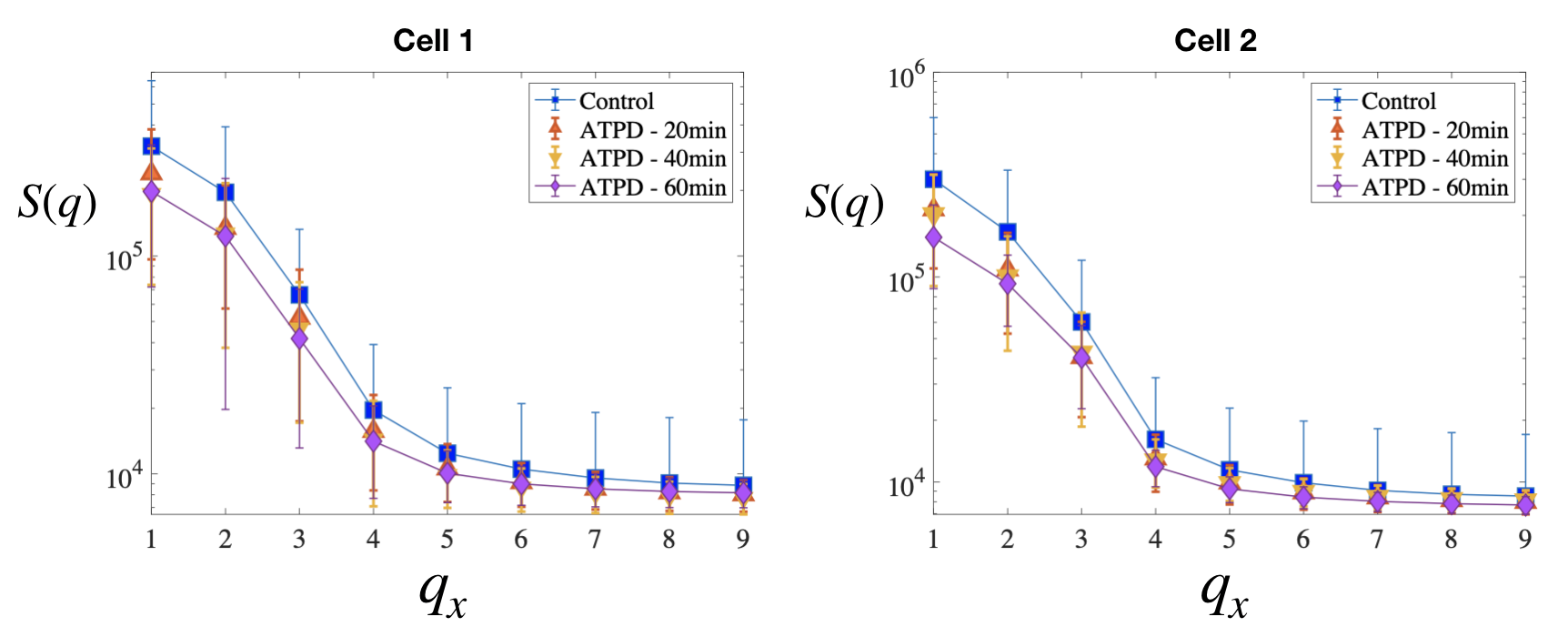}
\caption{The spectrum of height fluctuations, $S(\qq)$,
  as a function of $q$ in the horizontal direction for the two cells shown
  in Fig. 2 of the main paper.
  Different symbols denote different duration of ATP-depeltition. 
\label{fig:Sq}
}
\end{figure*}
\label{sec:Amodel}
We calculate the entropy production rate for an elementary model of
active membrane~\cite{prost1996shape}.  
Usually, flickering data is interpreted with a  model~\cite{muller2006biological,
  phillips2012physical,Safran2018statistical}
where the membrane is described by a height field $h(x,y)$  above
the $x$--$y$ plane in Monge gauge described by the Hamiltonian
\begin{equation}
  \mH[h] = \int \dx \dy \left[ \frac{\gamma}{2}\vert \grad h \vert^2 +
  \frac{\kappa}{2}\left(\lap  h\right)^2 \right] \/,
  \label{eq:Hamil}
\end{equation}
in equilibrium at temperature $T$.
Here $\gamma$ is the surface tension and $\kappa$ is the bending rigidity of the
membrane.
As the fluctuations of $h$ are small, the model uses a small deformation approximation
of curvature.
We shall call this the Gaussian model. 
The Hamiltonian is diagonal in Fourier space. 
In the same spirit as Refs.~\cite{prost1996shape, turlier2016equilibrium,
  turlier2019unveiling}
we write an active version of this model
by writing an equation of motion for each Fourier mode $\hhat(\qq,t)$
coupled to an
Ornstein--Uhlenbeck noise ~\cite{chandrasekhar1943stochastic} $\lambda(\qq,t)$  as,
\begin{subequations}
  \begin{align}
\delt{\hhat(\qq,t)} 
     +\frac{1}{\tau(\qq)}\; \hhat(\qq,t)+b\;\lambda(\qq,t) &=\sqrt{2D}\;\zeta(\qq,t)   \label{eq:dhdt}\\
  \delt{\lambda(\qq,t)} +\frac{1}{\tauA}\;\lambda(\qq) &= 
   \sqrt{2 \DA}\;\zetaA(\qq,t),\/,
   \label{eq:dldt}\\
 \text{where}\quad \quad \tau(\qq) &= \frac{4 \eta  q}{\kappa  q^4+\gamma q^2 }\/, \label{eq:tau}\\
  \bra{\zeta(\qq,t)\zeta(-\qq,s)} &= 2 D(\qq) \delta(t-s)\/,\label{eq:zeta}\\ 
  \bra{\zetaA(\qq,t)\zetaA(-\qq,s)} &= \delta(t-s)\/,\quad\text{and}\label{eq:zetaA}\\
    D(\qq) &= \frac{\kB T}{4 \eta q} \label{eq:D}
  \end{align}
  \label{eq:evol}
\end{subequations}
The membrane is assumed to undergo a stochastically driven overdamped motion.
The viscosity of the fluid around the membrane is $\eta$ which is related to
the amplitude of the stochastic driving $\zeta$ by fluctuation-dissipation
theorem~\eq{eq:D}.
In addition, there is active noise, $\zetaA$ which is assumed to be
white-in-space and an Ornstein--Uhlenbeck noise~\eq{eq:dldt} in time with a
correlation time $\tauA$ and an amplitude $\DA$. 
This active noise models both active processes on the
membrane~\cite{prost1996shape}, e.g., opening of ion channels, endo and
exo cytosis etc, and driving by the cytoskelton~\cite{turlier2016equilibrium}.
In \eq{eq:dhdt}, we use the parameter $b$ (of dimension $s^{-1}$) to determine
the strength of the coupling of the membrane to the active fluctuations.
Note that there are other models of time-correlated noises used to model active
fluctuations in living systems.
In Refs.~\cite{turlier2016equilibrium, turlier2019unveiling,prost1996shape},
the active noise is modelled using a random telegraph
process~\cite{gardiner1985handbook}, which leads to the same exponential
correlation in time for the noise as the Ornstein-Uhlenbeck process.
In Refs.~\cite{dabelow2019irreversibility,dabelow2021irreversibility,
  dabelow2021irreversible} the Ornstein--Uhlenbeck model itself is used.

For simplicity, from here onwards, we drop the $\qq$ dependence of
$\zetaA$, $\tauA$, and $\DA$.
As we have ignored nonlinearities in the model for the membrane
every $\hhat(\qq)$ is independent of every other $\hhat(\qq)$.
Consequently, the total entropy production rate is a sum over all $\qq$,
\begin{equation}
  \sigma = \sum \sigma_{\qq}\/.
  \label{eq:sum}
  \end{equation}
The individual terms in the sum  corresponding to the entropy production rate
for Eqs.~\ref{eq:dhdt} and \ref{eq:dldt} can be calculated
analytically~\cite{seifert2005entropy,ComplexL}, by mapping to a
well--studied model in stochastic
thermodynamics~\cite{Pal,manikandan2018exact,manikandan2021quantitative}.
We obtain,
\begin{subequations}
  \begin{align}
  \sigma_{\qq} &= 2 \kB\;\frac{b^2 \DA \tau(\qq) \tauA}{D (\tau(\qq) + \tauA)}
  \label{eq:sigmaq1} \\
  &= \frac{2\kB}{\tauA(\qq)}\left[
    \frac{S(\qq)\big\vert_{\rm A} }{S(\qq)\big\vert_{\rm Eq} }
    -1 \right], 
  \label{eq:sigmaq2}
  \end{align}
  \label{eq:sigmaq}
  \end{subequations}
In \eq{eq:sigmaq2} the subscript ``A'' denotes the variance being calculated
for an active membrane ($b\neq 0$) whereas the subscript ``Eq'' denotes the
variance being calculated for membrane in thermal equilibrium, i.e., $b = 0$.
They can be computed from the knowledge of the stationary distribution of
$\hhat$ and $\lambda$  as \cite{Pal,BGEXPT}
\begin{equation}
  S(\qq)\big\vert_{\rm A} = \avg{\hhat^*(\qq)\hhat(\qq)}
  = D\tau(\qq)+ \frac{b^2 \DA \tau(\qq)^2  \tauA^2}{\tau +\tauA}.
\end{equation}
We remark that the entropy production rate remains finite in the white noise
limit of the Ornstein--Uhlenbeck process. 

Notice that, though the model provides an exact analytical
expression~\eq{eq:sigmaq1} and an accessible form~\eq{eq:sigmaq2}
for the entropy production rate, applying it on the experimental data is
rather challenging.
This is because, we do not have access to either the active time-scale
$\tauA$ nor the equilibrium fluctuations of the membrane
$S(\qq)\big\vert_{\rm Eq}$.
Furthermore, in reality, the mechanical properties of the membrane such as
$\sigma$ and $\kappa$ could also change as a function of ATP depletion, and
the model ignores any such possibility~\cite{Activesolids,EffectiveTension}. 
Nevertheless, assuming the model is representative of the actual membrane
dynamics observed in the experiments, 
the following qualitative features can be deduced.\\
\begin{enumerate}
    \item The entropy production rate vanishes in the $\DA \rightarrow 0$ limit.
    \item The entropy production rate must be positive. Hence,
    \begin{align}
       S(\qq)\big\vert_{\rm A}  \geq S(\qq)\big\vert_{\rm Eq}
    \end{align}
    Assuming that the ATP depletion takes the cell membranes close to the
    equilibrium state, we can further argue that
    $ S(\qq)\big\vert_{\rm A} > S(\qq)\big\vert_{\rm ATPD}$
    where the subscript stands for the ATP depleted cell membrane.
    \item The summation in \eq{eq:sum} should converge. Hence,
    \begin{align}
      S(\qq)\big\vert_{\rm A} \rightarrow S(\qq)\big\vert_{\rm Eq}
      \text{ for large $q$}.
    \end{align}
    As a corollary, we also obtain
    $S(\qq)\big\vert_{\rm A} \rightarrow S(\qq)\big\vert_{\rm ATPD}$ for large $q$.
\end{enumerate}

Indeed, we find that fluctuations of the cell membranes we studied
experimentally exhibit the qualitative features predicted by this model.
We summarize our findings in \Fig{fig:Sq} for the two cell samples in
Fig. 2 of the main paper.
For simplicity, we have only shown $S(\qq)\big\vert$ for
$\qq[0,i]\equiv q_x$ with $1\leq i\leq 9$, and we find that it decreases as a
function of ATP depletion in time for a fixed $\qq$.


%


\end{document}